\documentclass[aps,prl,preprint,groupedaddress,preprintnumbers]{revtex4}

\usepackage{epsfig}
\usepackage{amssymb}
\usepackage{times}

\begin{document}

\newcommand{\re}{\mathop{\mathrm{Re}}}
\newcommand{\im}{\mathop{\mathrm{Im}}}
\newcommand{\I}{\mathop{\mathrm{i}}}
\newcommand{\D}{\mathop{\mathrm{d}}}
\newcommand{\E}{\mathop{\mathrm{e}}}

\def\lambar{\lambda \hspace*{-5pt}{\rule [5pt]{4pt}{0.3pt}} \hspace*{1pt}}

\preprint{\Large DESY 21-036}


\title{
Observation of harmonic lasing in the Angstrom regime at European XFEL}



\author{E.A.~Schneidmiller}
\email[]{evgeny.schneidmiller@desy.de}
\author{F.~Brinker}
\author{W.~Decking}
\author{L.~Froehlich}
\author{M.~Guetg}
\author{D.~Noelle}
\author{M.~Scholz}
\author{M.V.~Yurkov}
\author{I.~Zagorodnov}
\affiliation{Deutsches Elektronen-Synchrotron (DESY),
Hamburg, Germany}

\author{G.~Geloni}
\author{N.~Gerasimova}
\author{J.~Gruenert}
\author{J.~Laksman}
\author{J.~Liu}
\author{S.~Karabekyan}
\author{N.~Kujala}
\author{Th.~Maltezopoulos}
\author{I.~Petrov}
\author{L.~Samoylova}
\author{S.~Serkez}
\author{H.~Sinn}
\author{F.~Wolff-Fabris}

\affiliation{European XFEL GmbH,
Schenefeld, Germany}

\begin{abstract}
Harmonic lasing provides an opportunity to extend the photon energy range of existing and planned X-ray FEL
user facilities. Contrary to nonlinear harmonic generation, harmonic lasing can generate a much more intense,
stable, and narrow-band FEL beam. Another interesting application is Harmonic Lasing
Self-Seeding (HLSS) that allows to improve the longitudinal coherence and spectral power of a
Self-Amplified Spontaneous Emission (SASE) FEL.
This concept was tested at FLASH in the range of 4.5 - 15 nm and at PAL XFEL at 1 nm.
In this paper we present recent results from the European XFEL where we successfully demonstrated
harmonic lasing at 5.9 Angstrom and 2.8 Angstrom. In the latter case we obtained both 3rd and
5th harmonic lasing and, for the first time, operated a harmonic lasing cascade (5th-3rd-1st harmonics of the undulator). These results pave the way for reaching very high photon energies, up to 100 keV.
\end{abstract}



{\small Published 19 March 2021 in Phys. Rev. Accel. Beams 24, 030701 

DOI:  https://journals.aps.org/prab/abstract/10.1103/PhysRevAccelBeams.24.030701

Copyright American Physical Society}

\bigskip

\bigskip

\bigskip

\bigskip

\maketitle


\section{INTRODUCTION}

Successful operation of X-ray free electron lasers (FELs)
down to the Angstrom regime opens up new horizons for photon science. Even shorter
wavelengths are requested by the scientific community.
A possible way to extend the photon energy range of high-gain X-ray FELs is to use harmonic
lasing, which is
the FEL instability at an odd harmonic of the
planar undulator \cite{murphy,hg-2,kim-1,mcneil,sy-harm} developing independently from the lasing at the fundamental.
Another option, proposed in \cite{sy-harm}, is the possibility to improve spectral brightness of an X-ray FEL by
the combined lasing on a harmonic in the first part of the undulator (with an increased undulator parameter K)
and on the fundamental in the second part of the undulator. Later this concept was named Harmonic Lasing Self-Seeded
FEL (HLSS FEL) \cite{hlss}.

Harmonic lasing was originally proposed for FEL oscillators \cite{colson} and was tested experimentally in infrared
and visible wavelength
ranges. It was, however, not demonstrated in high-gain FELs and at a short wavelength until the
successful experiments \cite{prab-hlss-fl2} at the second branch of the soft X-ray FEL user facility
FLASH \cite{flash,fl2-njp} where the HLSS FEL operated in the wavelength range between 4.5 nm and 15 nm. Later,
the same operation mode was tested at PAL XFEL at 1 nm \cite{pal-hlss}. In this paper we report on recent
results from the European XFEL \cite{winni-xfel}, where we successfully demonstrated
operation of HLSS scheme at 5.9 Angstrom and at 2.8 Angstrom. Moreover, we observed for the first time the fifth
harmonic lasing in a high-gain FEL and operation of a harmonic lasing cascade.

\section{HARMONIC LASING}

Harmonic lasing in single-pass high-gain FELs \cite{murphy,hg-2,kim-1,mcneil,sy-harm} is the
amplification process, in a planar undulator, of higher odd harmonics developing independently of each
other (and of the fundamental) in the exponential gain regime.
The most attractive feature of the saturated harmonic lasing
is that the spectral brightness (or brilliance)
of harmonics is comparable to that of the fundamental \cite{sy-harm}.
Indeed, a good estimate for the
saturation efficiency is $\lambda_{\mathrm{w}}/(h L_{\mathrm{sat},h})$, where $\lambda_{\mathrm{w}}$ is the
undulator period, $h$ is the harmonic number, and
$L_{\mathrm{sat},h}$ is the saturation length of the h-th harmonic.
At the same time, the relative rms bandwidth
has the same scaling. So, the reduction of power is compensated by the bandwidth reduction and the spectral
power remains the same.

Although known theoretically for a long time \cite{murphy,hg-2,kim-1,mcneil}, harmonic lasing in high-gain FELs
was not considered for practical applications in X-ray FELs.
The situation was changed after publication of Ref.~\cite{sy-harm} where it was concluded that
the harmonic lasing in X-ray FELs is much more robust than usually thought, and can be effectively used in
the existing and future X-ray FELs. In particular, the European XFEL 
can greatly outperform
the specifications \cite{euro-xfel-tdr}
in terms of the highest possible photon energy: it can reach 60-100 keV range for the
third harmonic lasing.
It was also shown \cite{cw} that one can keep sub-Angstrom range of operation of the European XFEL
after CW upgrade
of the accelerator, even with a reduction of the electron energy from 17.5 GeV to 7 GeV.
Another application of harmonic lasing is a possible upgrade of FLASH with the aim to
increase the photon energy
up to 1 keV \cite{fl-harm-las}.

\subsection{Harmonic Lasing Self-Seeded FEL}

The poor longitudinal coherence of SASE FELs stimulated theoretical and experimental efforts for its improvement.
Since external seeding seems to be difficult to realize in the X-ray regime, a so called self-seeding scheme has been proposed
\cite{ss-soft,ss-wake}.
There are alternative approaches for reducing bandwidth and increasing spectral brightness of X-ray FELs without using
optical elements. One of them was proposed in \cite{sy-harm}, and is based on the combined lasing on a harmonic in
the first part of the undulator
(with increased undulator parameter K)
and on the fundamental in the second part. In this way the second part of the undulator is seeded by a narrow-band signal,
generated via harmonic lasing in the first part.
This concept was named HLSS FEL \cite{hlss}. Note that a similar concept, called pSASE, was independently proposed in \cite{psase}.

\begin{figure}[tb]

\includegraphics[width=.7 \textwidth]{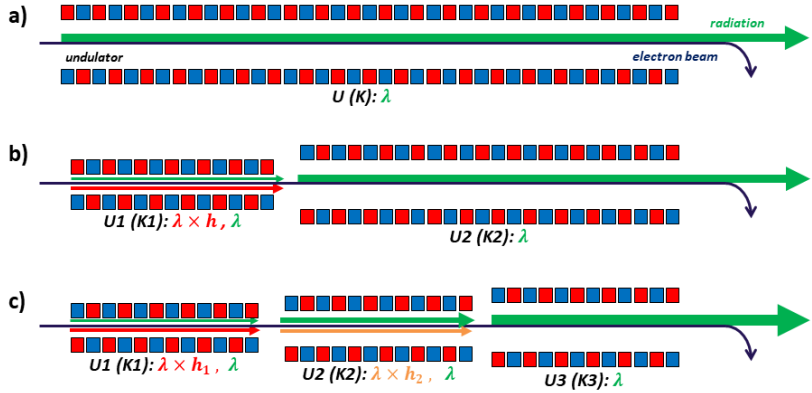}

\caption{\small Conceptual scheme of a SASE FEL (top), a Harmonic Lasing Self-Seeded FEL (middle) and a Harmonic Lasing Cascade (bottom). Green arrows correspond to the wavelength of interest $\lambda$, amplified through the whole undulator. Red and yellow arrows correspond to the fundamental wavelengths of the corresponding parts of the undulator, tuned to the subharmonic of $\lambda$.}

\label{hlss}
\end{figure}

Typically, gap-tunable undulators are planned to be used in  X-ray FEL facilities. If the maximal undulator parameter $K$ is
sufficiently large, the concept of harmonic lasing self-seeded FEL can be applied (see Fig.~\ref{hlss}b).
An undulator is divided into two
parts by setting two different undulator parameters such that the first part is tuned to a $h$-th sub-harmonic
of the second part, which is tuned to a wavelength of interest $\lambda$.
Harmonic lasing occurs in the exponential gain regime in the first part of the
undulator. The fundamental with wavelength $\lambda \times h$
in the first part stays well below
saturation. In the second part of the undulator, the fundamental is resonant to the
wavelength previously amplified as the harmonic. The amplification
process proceeds in the fundamental, up to saturation. In this case the
bandwidth is defined by the harmonic lasing but the saturation power is
still as high as in the reference case of lasing on the fundamental in the whole undulator, i.e.,
the spectral brightness increases.

The enhancement factor of the coherence length (or bandwidth reduction factor) that one obtains in HLSS FEL in comparison with a
reference case of lasing in the SASE FEL mode in the whole undulator, is given by \cite{hlss}:

\begin{equation}
R \simeq h \ \frac{\sqrt{L_{\mathrm{w}}^{(1)} L_{\mathrm{sat},h}}}{L_{\mathrm{sat},1}}
\label{R}
\end{equation}

\noindent Here $h$ is the harmonic number, $L_{\mathrm{sat},1}$ is the saturation length in the reference case of
the fundamental lasing with the lower K-value, $L_{\mathrm{w}}^{(1)}$ is the length of the first part of the undulator,
and $L_{\mathrm{sat},h}$ is the saturation length of harmonic lasing.

Despite the bandwidth reduction factor (\ref{R}) being significantly smaller than that of self-seeding schemes using
optical elements \cite{ss-soft,ss-wake}, the HLSS FEL scheme is very simple, robust, and it does not require any additional installations,
i.e., it can always be used in existing or
planned gap-tunable undulators with a sufficiently large range of accessible K-values.

\subsection{Harmonic Lasing Cascade}

Harmonic Lasing Cascade (see Fig.~\ref{hlss}c) is a conceptually simple generalization of HLSS towards increasing the number of undulator parts tuned to different subharmonics of the desirable frequency. This frequency is amplified through the whole chain being, for example, the fifth harmonic in the first part, the third harmonic in the second part, and the fundamental in the third part of the undulator. 

The scheme similar to that shown in Fig.~\ref{hlss}c can be applied when one installs a shorter period undulator behind the main undulator. For example, an installation of a short-period superconducting undulator after one of the two hard x-ray undulators of the European XFEL is being discussed \cite{sara}. The period of this linearly polarized device is chosen to be 15 mm, and the peak value of K is supposed to be 2.3 for the gap of 5 mm. For the electron energy of 17.5 GeV, the undulator may operated in the range 54  - 100 keV, 
In this case the main undulator can be divided into two parts with 5th and 3rd harmonic lasing, while the saturation is achieved at the fundamental wavelength of the superconducting undulator. For example, in 100 keV case, the first part of the main undulator is tuned to 20 keV, while the second part to 33 keV. Then 100 keV is amplified initially as the 5th harmonic, then as the third, and finally as the fundamental. 

The reason why we do not consider operation of the main undulator only on the fifth, or only on the third harmonic, is that one needs to keep the fundamental of that undulator well below saturation to avoid nonlinear harmonic generation. This can be done, in principle, with phase shifters \cite{mcneil} (for example, one can use $2\pi/3$ and $4\pi/3$ phase shifts in case of the 3rd harmonic). However, as it was shown in \cite{sy-harm,fl-harm-las}, this way of suppression is not very efficient in SASE case. Thus, dividing the undulator into two parts (with two different fundamental frequencies) makes the suppression task much easier \cite{cw, penn}.

\section{EXPERIMENTS AT THE EUROPEAN XFEL}

The European XFEL is an X-ray FEL user facility, based on a superconducting accelerator \cite{winni-xfel}. It
provides ultimately bright photon beams for user
experiments since 2017. Two hard X-ray undulators (SASE1 and SASE2) and one soft X-ray undulator (SASE3) are
presently in operation. Harmonic lasing experiments were performed in the SASE3 undulator
in April and in August 2019, and then continued in February 2020.

The electron beam energy in all cases was 14 GeV and the bunch charge was 250 pC. The SASE3 undulator consists of twenty one undulator segments, each 5 m long, and intersections containing quadrupoles and phase shifters between the segments
\cite{und}. The undulator has a period of 6.8 cm
and the K-values can be changed between 4 and 9 in the specified working range. Note that in order to
observe HLSS effect with the 5th harmonic we had to work outside of this standard range (K was set to 3.2).

\subsection{Third harmonic lasing at 2.1 keV}

\begin{figure}[t]
\includegraphics[width=.7\textwidth]{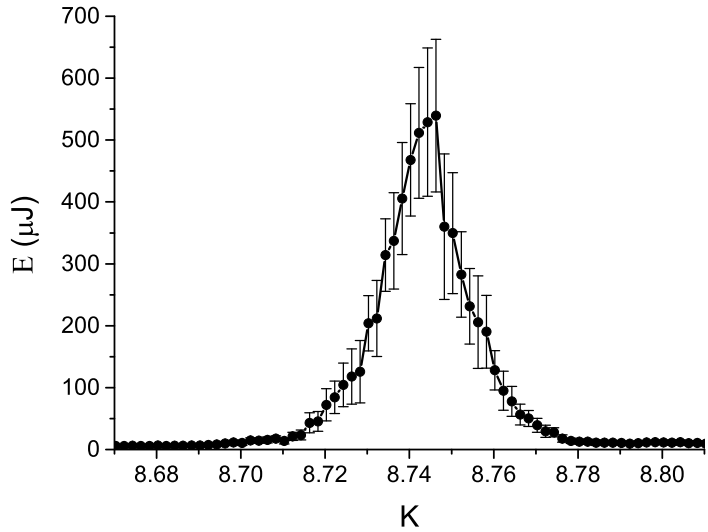}

\caption{\small Scan of the K-value of the first part of the undulator consisting of five undulator
segments, tuned to 700 eV. Mean pulse energy and its standard deviation are
measured after the second part of the undulator tuned to 2.1 keV. FEL operates in "5+7 segments HLSS" mode.
}

\label{K-scan}
\end{figure}

In April, 2019 we aimed at the 3rd harmonic lasing at 2.1 keV.
As a preparation step, we defined the number of undulator segments
to be tuned to the third subharmonic of 2.1 keV radiation that was our target photon energy. We set the
undulator K value to 8.74 and lased at 700 eV on the fundamental. 
In the first part of the undulator we have to avoid the nonlinear harmonic generation which means that we have
to stay three or more orders of magnitude below saturation. We have found that five undulator segments
satisfy this condition. Note that one could increase this length if one would use phase shifters to suppress the fundamental \cite{mcneil}. This, however, was not done during the experiments described in this paper due to the lack of time and a limited efficiency of this method \cite{sy-harm,fl-harm-las}. 

\begin{figure}[tb]
\includegraphics[width=.7\textwidth]{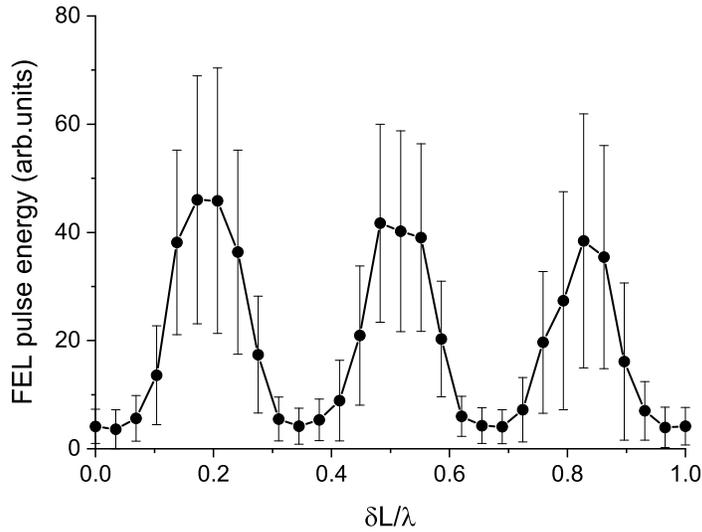}

\caption{\small Synchronous scan of phase shifters after first four undulator segments.
FEL pulse energy is measured after the second part of the undulator
tuned to 2.1 keV. Mean pulse energy and its standard deviation are plotted
versus a delay measured in units of the fundamental wavelength in the first part (tuned to 700 eV).
FEL operates in "5+6 segments HLSS" mode. 
}

\label{PS-scan}
\end{figure}

Then we set the undulator K-value to 4.91 and had a normal SASE operation at 2.1 keV
(wavelength 5.9 Angstrom). We chose twelve undulator
segments to stay at the onset of saturation. The pulse energy was 700 $\mu$J as measured by using an X-ray gas monitor
(XGM) \cite{xgm-1,xgm-2}. In the same way as it was done in the earlier experiments at FLASH \cite{prab-hlss-fl2},
we closed the 
gaps of the first five undulator segments one by one such that the undulator K became 8.74, and the resonance photon
energy was 700 eV (wavelength 17.7 Angstrom). In other words, we moved to the configuration, shown
on Fig.~\ref{hlss}, with $h=3$, five segments in the first part of the undulator and seven segments in
the second part.
As soon as K in each of the first five segments deviated from the original value of 4.91, we observed a strong reduction
of pulse energy, measured with an XGM. However, when K approached the value of 8.74, the pulse energy almost
recovered.

\begin{figure}[tb]
\includegraphics[width=.7\textwidth]{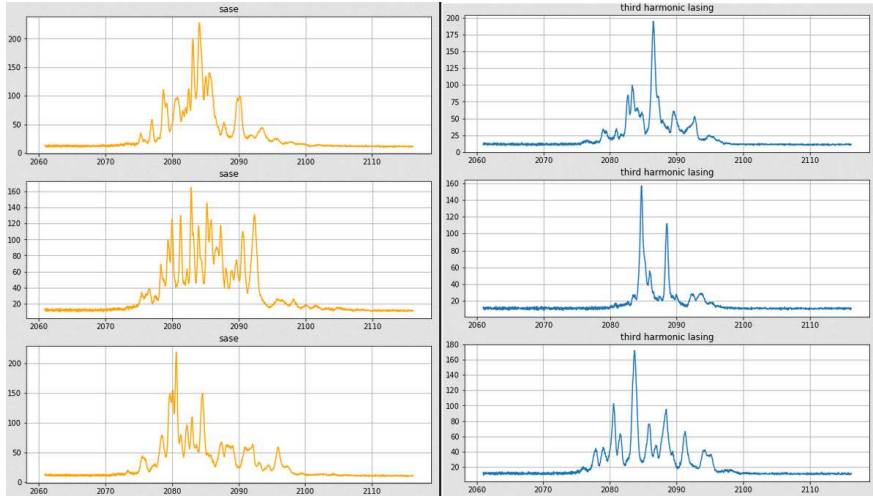}

\caption{\small Single-shot spectra for SASE (left column) and HLSS (right column).
Plots show spectral intensity (in arbitrary units) versus photon energy (in eV).
HLSS operates in "5+7 segments" configuration. One can see a reduction of number of spikes in the HLSS spectra.
}
\label{spikes}
\end{figure}

We performed a scan of the K-value of the first part of the undulator, tuned to 700 eV, and measured pulse energy
after the second part, tuned to 2.1 keV. The result is presented in
Fig.~\ref{K-scan} where one can see a sharp resonance. Taking into account the above mentioned
fact that lasing on the fundamental at 700 eV was too weak to produce nonlinear harmonics, we can conclude that
we have the 3rd harmonic lasing in the first part of the undulator at 2.1 keV that continues as lasing
on the fundamental in the second part.

An independent confirmation of harmonic lasing is the result of a scan
of phase shifters in the first part of the undulator, presented in Fig.~\ref{PS-scan}.
For this scan, we used six undulator segments in the second part in order to increase sensitivity of the
system by reducing saturation effects. We performed synchronous scans of phase shifters between the first five
undulator segments, over a range in which the fundamental radiation at $\lambda = 17.7$ Angstrom slips
over the electron beam by a single cycle (note that absolute value of phase in Fig.~\ref{PS-scan} is not calibrated, i.e. zero on the plot does not correspond to zero phase shift). If we seeded the second part of the undulator by the nonlinear
harmonic generation in the first part, the phase shifter scan would result in a single cycle. However, one
can clearly see three cycles in Fig.~\ref{PS-scan} which proves that we had harmonic lasing.

The purpose of the HLSS scheme is to increase FEL coherence time and to reduce bandwidth. From Equation
(\ref{R}) we estimate the coherence enhancement factor $R \simeq 2$ for our experimental conditions.
An increase of the
coherence time results in a corresponding reduction of the number of longitudinal modes with respect to the
SASE case. In the frequency domain we should then observe a reduction of the average number of spikes in single-shot
spectra. In an ideal case, we could also expect a reduction of the average bandwidth by a factor of two, to
the level of $0.1\%$ FWHM. Unfortunately, a reduction of the average bandwidth was not possible in our case,
since the bandwidth is strongly affected by the energy chirp of the electron bunches.

\begin{figure}[tb]
\includegraphics[width=.7\textwidth]{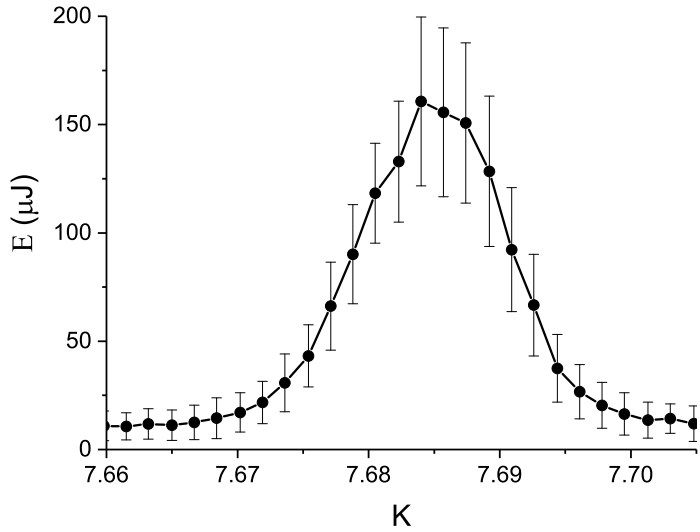}

\caption{\small Scan of the K-value of the first part of the undulator consisting of five undulator
segments, tuned to 900 eV. Mean pulse energy and its standard deviation are
measured after the second part of the undulator tuned to 4.5 keV. FEL operates in "5+11 segments HLSS" mode.
}

\label{K-scan-5th}
\end{figure}

To measure spectra, the SASE3 monochromator \cite{spectrometer} was used in spectrometer mode: the spectral
distribution of the soft X-rays was converted into visible light in a YAG:Ce crystal located in the focal
plane of the monochromator and was recorded by a CCD \cite{phot-diag}. To resolve spikes in single shot spectra, we reordered
the spectra in the third diffraction order with a 50 l/mm grating.

We took a series of spectra for the HLSS configuration (five segments at 700 eV and seven segments at 2.1 keV) as
well as for standard SASE with twelve segments at 2.1 keV. As expected, the average spectrum width was
dominated by the chirp and was the same
in both cases, $0.6\%$ FWHM. However, we could clearly observe a reduction of number of spikes in spectra.
This is illustrated by Fig.~\ref{spikes} where one can see three representative shots for each configuration.

\subsection{Third and fifth harmonic lasing at 4.5 keV}

\begin{figure}[tb]

\includegraphics[width=.7\textwidth]{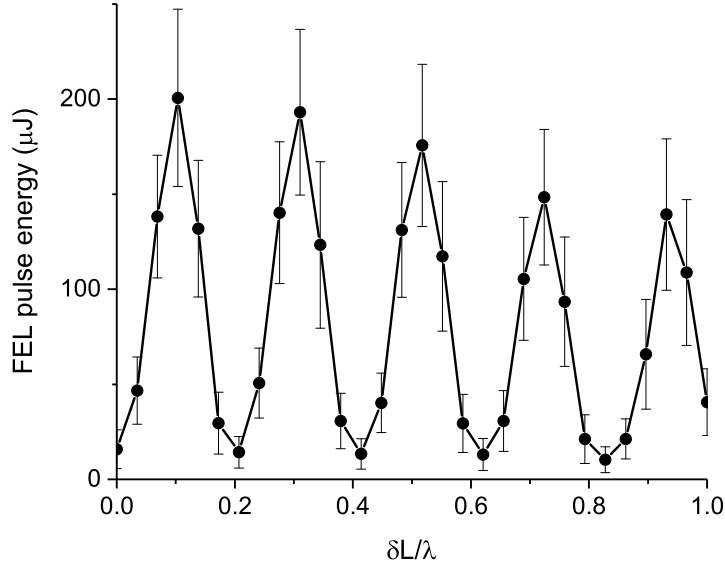}

\caption{\small Synchronous scan of phase shifters after first four undulator segments.
The FEL pulse energy is measured after the third part of the undulator
tuned to 4.5 keV. Mean pulse energy and its standard deviation are plotted
versus a delay measured in units of the fundamental wavelength in the first part (tuned to 900 eV).
The FEL operates as a harmonic lasing cascade (five segments at 900 eV, six segments at 1.5 keV, and six segments at 4.5 keV).
}

\label{PS-scan-5th}
\end{figure}

In August 2019 and in February 2020, we went to 4.5 keV lasing on the fundamental (2.8  Angstrom) and then tested the HLSS scheme
on the 3rd harmonic, by setting the first five segments to 1.5 keV. After that we set these segments to 900 eV and
obtained HLSS operation with the 5th harmonic seeding. In both cases the fundamental in the first part of
the undulator was well below saturation, so that there were no nonlinear harmonics. 
We performed a scan of the K-value of the first part of the undulator, tuned to 900 eV, and measured the pulse energy
after the second part, tuned to 4.5 keV. The result is presented in
Fig.~\ref{K-scan-5th}. Note that the spectrum at 4.5 keV can not be measured since the transmission of photon transport line has a cutoff at 3 keV.

Finally, we changed to a Harmonic Lasing Cascade configuration with the five first segments tuned to 900 eV, the following six segments to 1.5 keV, and the last six segments to 4.5 keV. In this case the spectral component at 4.5 keV is amplified as the fifth harmonic in the first part, then as the third harmonic in the second part with final amplification as the fundamental in the last part of the undulator. First two cascades are operated in the linear regime while in the last cascade the nonlinear regime is reached. Pulse energy was about 200 uJ for the cascade configuration as well as for SASE.
In Fig.~\ref{PS-scan-5th} we show the result of synchronous scan of phase shifters after first four undulator segments (tuned to 900 eV).
FEL pulse energy is measured at 4.5 keV (after the whole harmonic cascade) 
versus a delay measured in units of the fundamental wavelength in the first part. The scans in the second and third parts of the cascade show, respectively, three cycles (similar to Fig.~\ref{PS-scan}) and one cycle.    

\section{CONCLUSION AND OUTLOOK}

We observed, for the first time, harmonic lasing in the Angstrom regime, both in 
Harmonic Lasing Self-Seeded FEL mode and Harmonic Lasing Cascade mode. This is an important milestone on the way 
to lasing at high photon energies (up to ~100 keV) \cite{100keV} as well as to improving photon beam properties at low energies \cite{hlss}. Further steps in this direction will include tests of Harmonic Lasing Cascade, without lasing on the fundamental in SASE3, as well as demonstration of harmonic lasing in the hard x-ray undulators of the European XFEL (SASE1 or SASE2).

\section{ACKNOWLEDGMENT}

The authors would like to thank Wim Leemans for the careful reading of the manuscript and helpful comments.

\end{document}